\documentclass[aps,prl,twocolumn,superscriptaddress,longbibliography]{revtex4}

\usepackage[english]{babel} 
\usepackage[latin1]{inputenc}
\usepackage[usenames,dvipsnames]{color}
\usepackage{graphicx}
\usepackage{amsmath,amssymb,amsfonts}
\usepackage[colorlinks=true,linkcolor=blue,urlcolor=blue,citecolor=blue]{hyperref}
\usepackage[percent]{overpic}
\usepackage{physics}

\begin{document}

\title{Controlling energy storage crossing quantum phase transitions\\ in an integrable spin quantum battery}

\author{Riccardo Grazi}
\affiliation{Dipartimento di Fisica, Universit\`a di Genova, Via Dodecaneso 33, 16146, Genova, Italy}
\author{Daniel Sacco Shaikh}
\affiliation{Dipartimento di Fisica, Universit\`a di Genova, Via Dodecaneso 33, 16146, Genova, Italy}
\author{Maura Sassetti}
\affiliation{Dipartimento di Fisica, Universit\`a di Genova, Via Dodecaneso 33, 16146, Genova, Italy}
\affiliation{CNR-SPIN, Via Dodecaneso 33, 16146, Genova, Italy}
\author{Niccol\'o Traverso Ziani}
\affiliation{Dipartimento di Fisica, Universit\`a di Genova, Via Dodecaneso 33, 16146, Genova, Italy}
\affiliation{CNR-SPIN, Via Dodecaneso 33, 16146, Genova, Italy}
\author{Dario Ferraro}
\affiliation{Dipartimento di Fisica, Universit\`a di Genova, Via Dodecaneso 33, 16146, Genova, Italy}
\affiliation{CNR-SPIN, Via Dodecaneso 33, 16146, Genova, Italy}

\begin{abstract}
We investigate the performance of a one-dimensional dimerized XY chain as a spin quantum battery. Such integrable model shows a rich quantum phase diagram that emerges through a mapping of the spins onto auxiliary fermionic degrees of freedom. We consider a charging protocol relying on the double quench of an internal parameter, namely the strength of the dimerization, and address the energy stored in the systems. We observe three distinct regimes, depending on the time-scale characterizing the duration of the charging: a short-time regime related to the dynamics of the single dimers, a long-time regime related to the recurrence time of the system at finite size, and a thermodynamic limit time regime. In the latter, the energy stored is almost unaffected by the charging time and the precise values of the charging parameters, provided the quench crosses a quantum phase transition. Such a robust many-body effect, that characterizes also other models like the quantum Ising chain in a transverse field, as we prove analytically, can play a relevant role in the design of stable solid-state quantum batteries.
\end{abstract}

\maketitle

\emph{Introduction.-} While the first quantum revolution was driven by gedanken experiments devoted to the comprehension of the rules governing the microscopic world, during the second one, which is currently underway, these ideas are turned into reality~\cite{Jaeger_Book}. Accordingly, it is now possible to create, manipulate and measure quantum systems with astonishing technological follow-ups. In this direction, the impact of quantum technologies in the fields of communication and computation is already evident. New applications are also emerging~\cite{Benenti_Book}. Among them, the study of devices able to store and transfer energy exploiting purely quantum features, the so called quantum batteries (QBs), started ten years ago and is developing at a very fast pace~\cite{Campaioli18, Bhattacharjee21,   Campaioli23}. 

The platforms typically considered for such devices are based on collections of (pseudo)-spins~\cite{Alicki13, Binder15, Hu22, Gemme22} promoted from the ground to the excited states by means of coupling with external cavities playing the role of chargers~\cite{Ferraro18, Andolina19, Ghym23, Liu21, Shi22, Delmonte21, Quach22, Dou22, Erdman22, Gemme23, Crescente23, Yang23} or through direct interaction among them~\cite{Le18, Liu19, Zhang19, Rosa20, Rossini20, Zhao22}. 

In the latter configuration, usually dubbed spin-QBs, general theorems state that the range of interaction and the coordination of the lattice created by the spins play a crucial role in preventing a super-extensive scaling of the average charging power, namely the energy stored into the device divided to the time required to complete the charging~\cite{Campaioli17, Julia20, Gyhm22}. At the same time, it is known that the presence of interactions can lead to quantum phase transitions (QPTs) which, changing the nature of the ground state of the system, can have a major impact on the time evolution~\cite{qq1,qq2} and hence in increasing the energy stored in the QBs~\cite{Barra22}. The accurate investigation of these aspects is usually strongly limited by the fact that this kind of systems typically requires numerical treatments, such as exact diagonalization, which are very demanding in terms computational resources.

Nevertheless, a particular class of interacting spin systems admits an exact solution~\cite{franchini2017introduction}. The study of these integrable systems in the framework of QBs is at an early stage~\cite{Catalano23}, but can enable major achievements in QB physics, also in view of the fact that they can be simulated in platforms used for quantum computation and quantum simulation such as Rydberg atoms~\cite{adams2019rydberg, Browaeys20, Wu21}. 

The present Letter aims at filling this gap presenting the theoretical analysis of a spin-QB based on an integrable dimerized XY chain. This model presents a rich phase diagram~\cite{perk1975soluble} and maps, after a Jordan-Wigner transformation, onto the dimerized Kitaev chain~\cite{wakatsuki2014fermion}. The main reason for the choice of the model is that it allows a direct comparison between usual few-body QBs and many-body ones. Indeed, when the system is fully dimerized it can be described as a collection of independent dimers, while at the QPTs all correlation lengths diverge and the behavior becomes strongly collective. Differently from what usually proposed in literature, where the energy is provided by external classical or quantum chargers, the charging of the QB is here achieved by changing, for a finite time $\tau$, one of the parameters of the system, notably the intensity of the dimerization. This situation can be realized, for example, by approaching some of the sites of the chain in spite of the others.

The main result of our analysis is the identification of three distinct regimes for the energy storage as a function of the charging duration $\tau$. For short $\tau=\tau_s$ the stored energy reaches its maximum in correspondence of the parameter range of the full dimerization. For large $\tau=\tau_r$ of the order of the recurrence time of the system \cite{rossini2020dynamics} the energy stored is again maximal in correspondence of the regime of full dimerization. However, in this regime, signatures of the QPTs of the system also emerge. Even more interestingly, at intermediate times, long in comparison to $\tau_s$ but short in comparison to $\tau_r$, the energy stored is basically independent of both the charging time and the specific values of the charging Hamiltonian, provided that the quench crosses a phase boundary (PB). This regime, the only one which survives in the thermodynamic limit $N\rightarrow +\infty$, identifies a stable and robust working point for the many-body QB with relevant potential applications. The strong dependence of the energy stored in the thermodynamic regime on the quantum phase diagram is also manifest, as a substantial enhancement, in the paradigmatic case of QB based on the quantum Ising chain in a transverse field. This highlights the fact that the observed phenomenology is not peculiar of the model under investigation and can open interesting new perspectives in the framework of solid-state QBs.


\emph{Model.-} We consider a $N$ site one-dimensional XY chain of spin-$1/2$ constituents, coupled by means of an anisotropic and dimerized interaction~\cite{perk1975soluble}. We restrict to the case of even $N$ and zero temperature.
The Hamiltonian describing this system reads

\begin{eqnarray}
     H_B &=& -J \sum_{j=1}^N \left[1 - (-1)^j \delta\right]\times\nonumber\\
     &&\left[\left(\frac{1+\gamma}{2}\right) \sigma_j^x \sigma_{j+1}^x+\left(\frac{1-\gamma}{2}\right) \sigma_j^y \sigma_{j+1}^y\right], 
     \label{Tesi_Ham}
\end{eqnarray}
where the subscript $B$ indicates the fact that this system will be considered as a QB. 

Here $\sigma_j^\alpha$ (with $\alpha = x,y$) denote the conventional Pauli matrices corresponding to the $j$-th site spin. The parameter $J$ is the energy scale of the system, while $\gamma$ and $\delta$ characterize the strength of the anisotropy and the dimerization respectively. 

By considering for simplicity the even-parity sector of the model~\cite{Porta}, and adopting periodic boundary conditions $\sigma_{j+N}^\alpha \equiv \sigma_j^\alpha$, it is possible to diagonalize the above Hamiltonian by means of a standard Wigner-Jordan transformation into free spinless fermions \cite{wigner1928}. It is worth to mention that the fermionic version of the model is a dimerized Kitaev chain characterized by interesting topological features~\cite{wakatsuki2014fermion}. In terms of auxiliary fermionic annihilation operators $a_q$ and $b_q$, and assuming from now on $J=1$ as reference scale for the energies, we obtain (from now one we will consider $\hbar=1$)
\begin{equation}
    H_B = \sum_{q \in \Gamma} \left[\omega_{1,q} \left(a_q^\dag a_q - \frac{1}{2}\right) + \omega_{2,q} \left(b_q^\dag b_q - \frac{1}{2}\right)\right] \\
\end{equation}
where 
\begin{equation}
    \omega_{1/2,q} = 2 \sqrt{(1 \pm \gamma \delta)^2 \cos^2 \left(\frac{\pi q}{\mathcal{N}}\right) + (\delta \pm \gamma)^2 \sin^2 \left(\frac{\pi q}{\mathcal{N}}\right)},
\end{equation}
with $\mathcal{N}=N/2$ the number of dimers. 

Additional details concerning the diagonalization procedure are reported in the Supplementary Material~\cite{SM}.

\begin{figure}
    \centering
\includegraphics[scale=0.42]{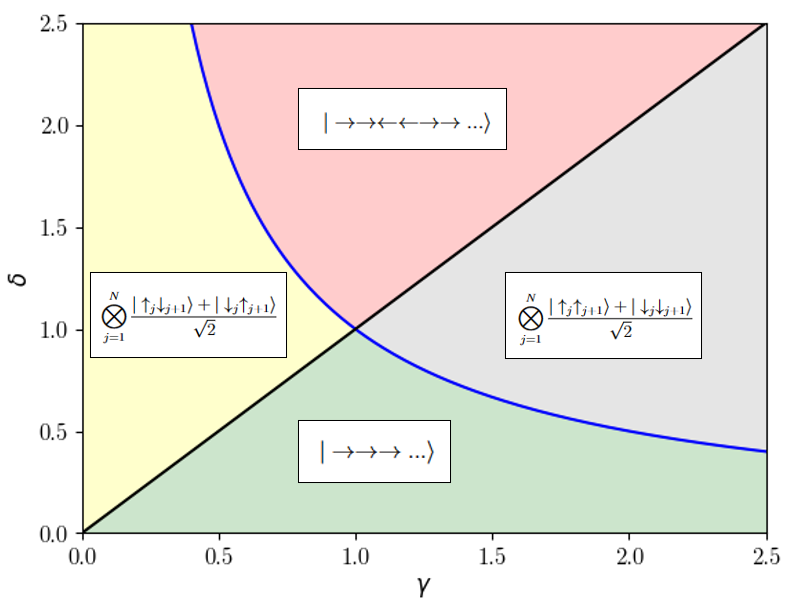}
    \caption{Phase diagram of the dimerized XY model for $\delta > 0$ and $\gamma > 0$. For every region we have indicated a schematic expression for the ground state of the system.}  
    \label{Figure1}
\end{figure}

\emph{Phase diagram.-} The zeros of the spectrum discussed above allow us to identify the PBs of the systems. They are given, in the thermodynamic limit $\mathcal{N}\rightarrow \infty$, by the conditions
$\gamma^2 \delta^2 = 1$ and $\delta^2 = \gamma^2$. The relative phase diagram in the region $\delta,\gamma > 0$ is shown in Fig.~\ref{Figure1}. The physical content of the ground state in these quantum phases can be appreciated recalling to the original spin description. To clarify this point one can examine the behavior of the system at representative points within each region of the diagram in Fig.~\ref{Figure1}.

\begin{itemize}
\item Region $(1)$ (green area).

For $\gamma = 1$ and $\delta = \frac{1}{2}$, the ground state, is given by a ferromagnetic state along the $\hat{x}$ direction, indicated schematically as e.g. $|\rightarrow\rightarrow\rightarrow...\rangle$.

\item Region $(2)$ (red area).

 For $\gamma = 1$ and in the limit $\delta \rightarrow \infty$, the ground state is given by a spin-1 antiferromagnetic state along the $\hat{x}$ direction schematically indicated by $|\rightarrow\rightarrow\leftarrow\leftarrow\rightarrow\rightarrow...\rangle$.

\item Region $(3)$ (grey area).

The case where $\gamma\rightarrow +\infty$ while maintaining $\delta = \frac{1}{2}$ results in a ground state given by the tensor product of dimers with spins aligned along the $\hat{z}$ direction, indicated by $
    \bigotimes_{j=1}^N \frac{|\uparrow_j \uparrow_{j+1}\rangle + |\downarrow_j \downarrow_{j+1}\rangle}{\sqrt{2}}.$

\item Region $(4)$ (yellow area).

Finally for this region we consider the point $\gamma=1/4$ and $\delta=1/2$. The ground state here is given by the tensor product of dimers with spins anti-aligned along the $\hat{z}$ direction, indicated by
$    \bigotimes_{j=1}^N \frac{|\uparrow_j \downarrow_{j+1}\rangle + |\downarrow_j \uparrow_{j+1}\rangle}{\sqrt{2}}.$

\end{itemize}
Notice that the line at $\delta=1$ corresponds to a fully dimerized chain, where many-body effects are absent.

The role of these quantum phases in the charging performance of the system as a QB will be discussed in the following. 

\emph{Quantum battery setup.-} We now analyze the behavior of the model when employed as a QB. To do so, we need to define the initial state of the QB and to introduce a charging protocol. 
As initial state, we chose the vacuum of both the $a_q$ and the $b_q$ fermions, namely the state $|0\rangle$ satisfying $
    a_q |0\rangle = b_q|0\rangle = 0 ~~ \forall q \in \Gamma$ (empty QB).
Our charging protocol consists in the double quench (on and off) of the strength of the dimerization: keeping fixed the anisotropy $\gamma$, we vary $\delta$ as $\delta(t)=\delta_{0}+\delta_{1}\theta(t)\theta(\tau-t)$,
where $\delta_0$ and $\delta_1$ are constant positive parameters, $\tau>0$ is the time associated to the quench and $\theta(x)$ is the Heaviside step function. To realize this practice one needs to engineer two distinct XY spin-chains, for the even and odd sites respectively, and shift one with respect to the other. Such approach seems within reach in view the current astonishing level of control of platforms for quantum simulations, such as the ones based on solid-state devices~\cite{Kampen05}, trapped ions~\cite{Senko15} and in particular Rydberg atoms~\cite{Barredo15, Crescimanna23}.

In the following, we will indicate with $H'_{B}$ the Hamiltonian in the time interval $t\in [0, \tau]$ (with $\delta=\delta_{0}+\delta_{1}$) and $H_{B}$ otherwise. Obviously also $H'_{B}$ can be diagonalized in exactly the same way as before and its diagonal form reads 
\begin{equation}
    H'_{B} = \sum_{q \in \Gamma} \left[\omega'_{1,q} \left(c_q^\dag c_q - \frac{1}{2}\right) + \omega'_{2,q} \left(d_q^\dag d_q - \frac{1}{2}\right)\right] 
\end{equation}
where $c_q$ and $d_q$ are new fermionic annihilation operators and $\omega'_{1,2}$ are the eigenvalues of $H'_{B}$. Introducing the short notations $\textbf{a}_{q}^{\dagger}= (a^{\dagger}_q, b^{\dagger}_q, a_{\mathcal{N}-q}, b_{\mathcal{N}-q})$ and $\textbf{c}_{q}^{\dagger}= (c^{\dagger}_q, d^{\dagger}_q, c_{\mathcal{N}-q}, d_{\mathcal{N}-q})$ the mapping between the two sets of operators is obtained through the transformation $\textbf{c}_{q} = \mathcal{V}_{q}^{-1} \mathcal{U}_{q} ~\textbf{a}_{q} \equiv \mathcal{M}_{q}\textbf{a}_{q}$,
where $\mathcal{U}_{q}$ ($\mathcal{V}_{q}$) is the 4x4 matrix having the eigenvectors of $H_B$ ($H'_B$) as columns. The explicit form of the matrix $\mathcal{U}_{q}$ has been obtained numerically and is too lengthy to be reported explicitly. 

The energy stored in the QB for a given time $t$ in the interval $\left[0, \tau\right]$ is then~\cite{Andolina18}
\begin{equation}
    \Delta E (t) = \langle \Psi(t)|H_{B}|\Psi(t)\rangle - \langle \Psi(0)|H_{B}|\Psi(0)\rangle \label{E_imm}
\end{equation}
where $
    |\Psi(t)\rangle = e^{-i H'_{B} t} |\Psi(0)\rangle
    \label{46}
$, with $|\Psi(0)\rangle\equiv |0\rangle$ the state of the system at $t=0$. Notice that here we are assuming a closed evolution for the system neglecting possible relaxation effects due to the external environment~\cite{Petruccione_book}.

We find
\begin{widetext}
\begin{equation}
\begin{aligned}
    \Delta E(t) &=  \sum_{q \in \Gamma_{}} \sum_{s_1}\omega_{s_1,q} n_{s_1,q}(t)=\sum_{q \in \Gamma_{}} \sum_{s,s_{1},s_{2},s_3 = 1,2} 2\omega_{s_1,q} \bigg\{\mathcal{M}_{s+2,s_{1}; q}\mathcal{M}^*_{s_{2}+2,s_{1}; q}\mathcal{M}^*_{s+2,s_3 + 2; q}\mathcal{M}_{s_{2}+2,s_3 + 2;q}~ \cos\left[\left(\omega'_{s,q} - \omega'_{s_{2},q}\right)t\right] + \\&+ \mathcal{M}_{s+2,s_{1};q}\mathcal{M}^*_{s_{2},s_{1};q}\mathcal{M}^*_{s+2,s_3 + 2;q}\mathcal{M}_{s_{2},s_3 + 2;q}~ \cos\left[\left(\omega'_{s,q} + \omega'_{s_{2},q}\right)t\right] \bigg\}, \label{DeltaE}
\end{aligned}
\end{equation}
\end{widetext}
with $n_{1,q}(t)=\bra{\psi(t)}a^\dag_q a_q\ket{\psi(t)}$, $n_{2,q}(t)=\bra{\psi(t)}b^\dag_q b_q\ket{\psi(t)}$, and $\mathcal{M}_{\eta,\eta';q}$ indicating the entries of the matrix $\mathcal{M}_{q}$. A detailed analysis of $n_{1,q}(t)$ and $n_{2,q}(t)$, is provided in~\cite{SM}, while in the following we will discuss the main features associated to Eq.~(\ref{DeltaE}).

\begin{figure}
    \centering   
    \includegraphics[scale=0.18]{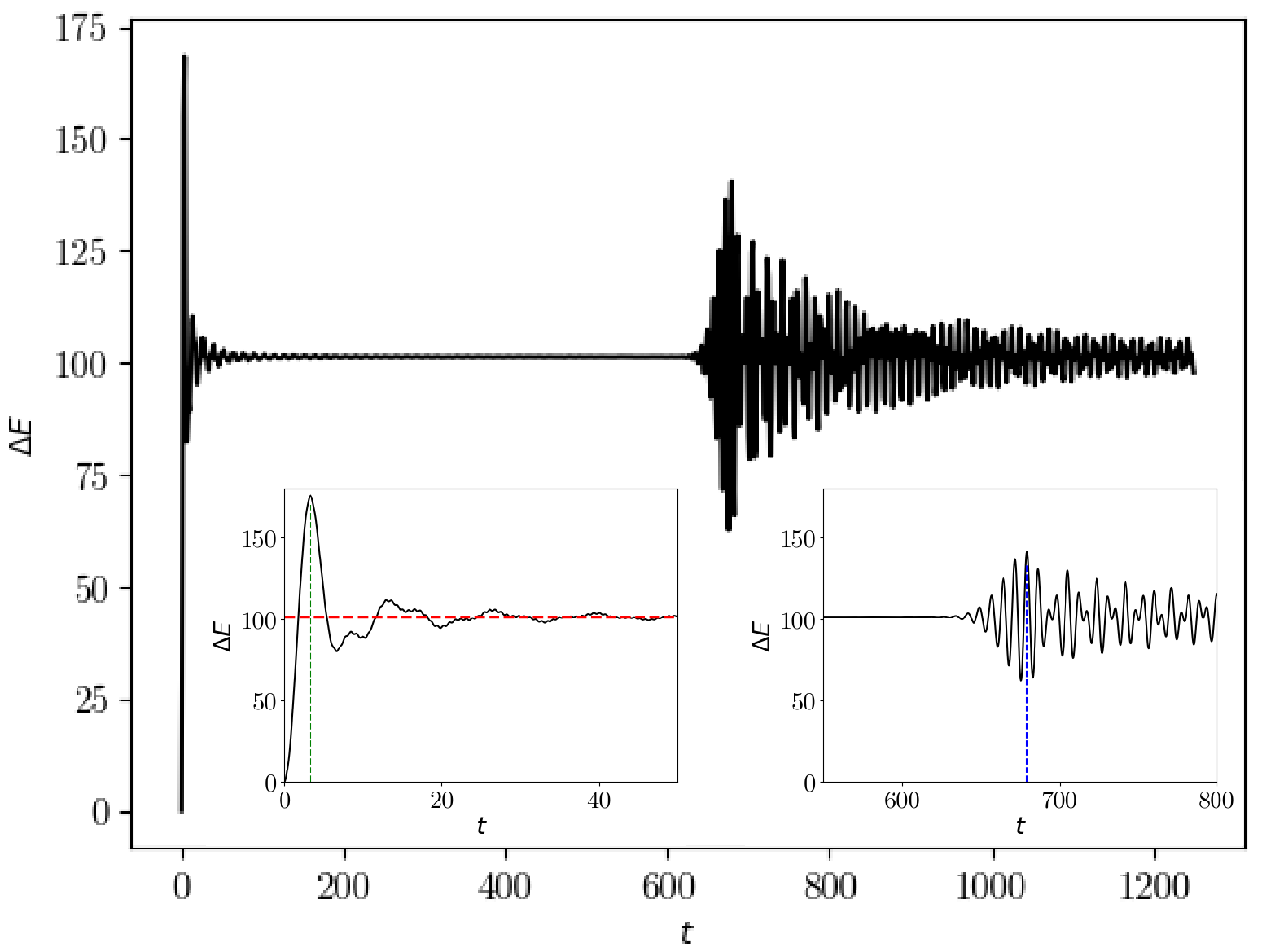}
    \caption{Energy $\Delta E$ stored in the QB as a function of time for $\delta_0 = 0.3$, $\delta_1 = 0.6$, $\gamma = 1.25$ and $\mathcal{N} = 300$. Left inset: zoom for $t \in \left[0, 50\right]$. The vertical green dotted line indicates the time $\tau_s$ at which the first maximum of the energy stored is achieved, while the horizontal red dotted line indicates the asymptotic limit of the maximum energy stored per dimer. Right inset: zoom for $t \in \left[600, 800\right]$. The vertical blue dotted line indicates the time $\tau_r$ at which the maximum of the energy stored during the recurrence is achieved.}
    \label{Figure2}
\end{figure}

\begin{figure}
    \centering   
    \includegraphics[scale=0.45]{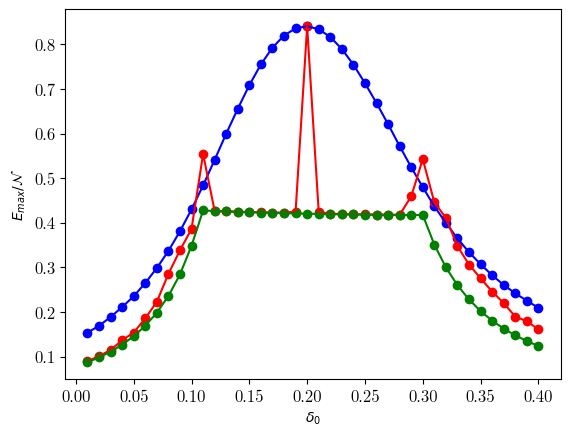}
    \caption{Plot of $E_{max}/\mathcal{N}$ as a function of $\delta_0$ for $\delta_1 = 0.8$, $\gamma = 1.1$ and $\mathcal{N} = 300$ in the short-time (blue), recurrence (red) and asymptotic (green) regimes.}
    \label{Figure3}
\end{figure}

The plot of $\Delta E(t)$ as a function of time for a representative choice of parameters is shown in Fig.~(\ref{Figure2}) where it is possible to identify three regimes. The first is a short-time one, characterized by strong oscillations (see left inset). Here, the maximal stored energy is $E_{max}^s$ and is achieved for $t=\tau_s$. The second one, which we call the thermodynamic regime, is characterized by a very flat behavior of $\Delta E(t)$ (strongly suppressed oscillations). The value of the energy stored in this region is indicated as $E_{max}^{\infty}$. This region extends up to the onset of the third regime, occurring at a time $t\sim \tau_r$ (the $r$ standing for recurrence), where oscillations dramatically emerges again (see right inset) as a consequence of the discreteness of the energy levels in finite systems~\cite{rossini2020dynamics}. Here $\tau_{r}$ grows linearly with the size of the system~\cite{SM} and the maximal energy is dubbed $E_{max}^r$. 
It is worth to mention that the features discussed above emerge for all tested charging protocols and system sizes whenever $N\geq 10$~\cite{SM}. Moreover, energies $E_{max}^s$,  $E_{max}^r$, and  $E_{max}^\infty$ grow linearly with $\mathcal{N}$ (and consequently with $N$) and show very different properties as a function of the parameters of the system. Their characterization is presented in Fig.~(\ref{Figure3}), where they are plotted as a function of the dimerization $\delta_0$ at fixed $\delta_1$ and $\gamma$. We observe that $E_{max}^s$, in blue, has its maximum in correspondence of $\delta_0+\delta_1=1$, namely when the Hamiltonian $H'_B$ describes a collection of independent dimers. The energy $E_{max}^r$, in red, again shows its maximum for $\delta_0+\delta_1=1$. The QPTs of the model are however manifest in this case as spikes located in correspondence of the values of $\delta_0$ for which $H'_B$ is critical. Finally, $E_{max}^\infty$, in green, shows a steady increase up to the first QPT, followed by a plateau, and hence a decrease after the second critical point. 

Two comments are in order at this point. First of all, we notice that, in spite of comparable stored energy, due to the discussed hierarchy in the charging times the three processes may be characterized by very different average charging power (stored energy divided by the charging time). Moreover, the peak of $E^{s}_{max}=E^{r}_{max}$ is greater with respect to the one of $E^{\infty}_{max}$ (roughly $2$ times in the considered range of parameters). However, reaching the latter doesn't require a careful control of the charging time due to its stability with an evident advantage for actual experimental implementations.

\begin{figure}
    \centering   
    \includegraphics[scale=0.45]{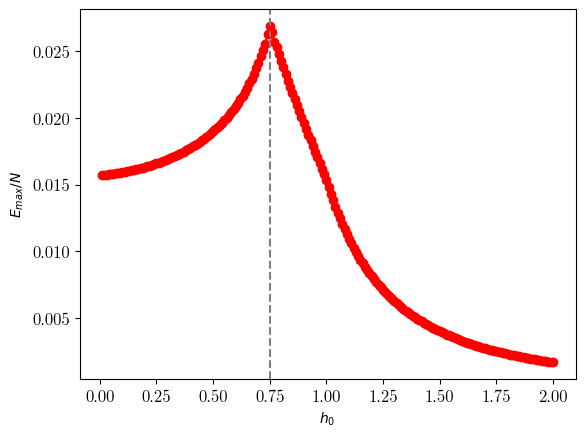}
    \caption{Plot of $E^{\infty}_{max}/N$ as a function of $h_0$ for the Ising model in the asymptotic regime, fixing $h_1 = 0.25$ and $N = 600$ spins. The gray dotted line indicates the value of $h_0$ such that $h_0 + h_1 = 1$.}
    \label{Figure4}
\end{figure}

This stability of the stored energy $E_{max}^\infty$ with respect to charging times and charging parameters, provided the charging protocol crosses one PB represents a major result of the present article.
To prove its generality, we also test it in the case of another model, namely the quantum Ising chain in a transverse field. This case is described by the Hamiltonian
\begin{equation}
    H_I(h) = \frac{1}{2} \sum_{j=1}^N \left[\sigma_j^x \sigma_{j+1}^x+h \sigma_j^z\right]. \label{XY_Ham}
\end{equation}
Here, other than already introduced notations, $h$ is the transverse field. As a Hamiltonian of the battery we adopt $H_I(h_0)$ and as a charging Hamiltonian $H'_B$ we use $H_I(h_0+h_1)$ in a spirit similar to the one discussed in Ref.~\cite{Le18}. The analytical form of the energy stored as a function of time is reported in~\cite{SM} and is qualitatively similar to Fig.~(\ref{Figure2}). The energy $E^{\infty}_{max}$ is then plotted in Fig.~(\ref{Figure4}). Despite the smaller absolute value of the stored energy with respect to the dimerized XY chain, it is clear that a strong enhancement of the stored energy is achieved also in this case for $h_0+h_1=1$, namely when $H'_B$ is critical.\\


\emph{Conclusions.-} 
We have investigated a spin quantum battery based on a dimerized XY chain, considering a charging protocol based on the double quench of the intensity of the dimerization. The integrability of this model allowed us to study the stored energy in regimes inaccessible for numerical methods such as exact diagonalization. We identify three working regimes. The first one is a short-time regime, where the most favorable charging protocol involves the evolution through a Hamiltonian that is factorized in disconnected dimers. The second is a regime occurring at long times, of the order of the recurrence time of the system. Also at this scale, the more favorable charging involves again the fully dimerized Hamiltonian, but signatures of the quantum phase transitions emerge. Finally, we have discussed a thermodynamic regime where, strikingly, the stored energy is very stable with respect to both the charging time and the charging parameters, as long as the charging Hamiltonian and the battery Hamiltonian are separated in phase space by a single critical line. This observation underlines the importance of the reorganization of the energy spectrum in the different phases of a spin chain to improve its performance, as also discussed in the framework of quantum thermal machines~\cite{Campisi16, Williamson24}, and does introduce a new paradigm for many-body quantum batteries. Finally, to further substantiate our findings, we have demonstrated that a spin quantum battery based on the Ising model in a transverse field also shows a remarkable dependence on the quantum phase transition of the model, in the form of an enhancement.

\begin{acknowledgments}

\emph{Acknowledgments.-} Authors would like to thank G. M. Andolina, F. Campaioli, A. Crescente and G. Gemme for useful discussions. N.T.Z. acknowledges the funding through the NextGenerationEu Curiosity Driven
Project "Understanding even-odd criticality". N.T.Z. and M.S. acknowledge the funding
through the "Non-reciprocal supercurrent and topological transitions in hybrid Nb- InSb
nanoflags" project (Prot. 2022PH852L) in the framework of PRIN 2022 initiative of the
Italian Ministry of University (MUR) for the National Research Program (PNR). 

D.F. acknowledges the contribution
of the European Union-NextGenerationEU through the
"Quantum Busses for Coherent Energy Transfer" (QUBERT) project, in the framework of the Curiosity Driven
2021 initiative of the University of Genova and through
the "Solid State Quantum Batteries: Characterization
and Optimization" (SoS-QuBa) project (Prot. 2022XK5CPX), in the framework of the PRIN 2022 initiative of the Italian Ministry
of University (MUR) for the National Research Program
(PNR).
\end{acknowledgments}


\end{document}


\begin{center}
    \textbf{SM1: Diagonalization of the dimerized XY Hamiltonian}
\end{center}
The Hamiltonian
\begin{SMEquations}
    H_B = -J \sum_{j=1}^N \left[1 - (-1)^j \delta\right]\left[\left(\frac{1+\gamma}{2}\right) \sigma_j^x \sigma_{j+1}^x+\left(\frac{1-\gamma}{2}\right) \sigma_j^y \sigma_{j+1}^y\right]
\end{SMEquations}
can be written in terms of ladder spin operators $\sigma^{\pm}_{j} \equiv \left(\sigma_{j}^x \pm i \sigma_{j}^y\right)/2$ as follows
\begin{SMEquations}
    H_B = -\frac{J}{2} \sum_{j=1}^N \left[1 - (-1)^j \delta\right]\left[(\sigma_j^+ \sigma_{j+1}^+ + \sigma_j^- \sigma_{j+1}^-)\gamma + \sigma_j^+ \sigma_{j+1}^- + \sigma_j^- \sigma_{j+1}^+ \right]. \label{HB_Ladder}
\end{SMEquations}
\noindent The standard way to diagonalize the Hamiltonian \eqref{HB_Ladder} consists in mapping the $\sigma^{\pm}$ operators into two auxiliary fermionic operators, indicated with $A_j$ and $B_j$. They are defined on the even and odd sites respectively and together form $\mathcal{N}=N/2$ dimers. The Jordan-Wigner transformations can be written as~\cite{franchini2017introduction}
\begin{equation}
\begin{aligned}
&\sigma_{j=even}^+ &=& \exp\{-i \pi \sum_{l<j}\left(A^\dag_l A_l + B^\dag_l B_l\right) \} A_j, \nonumber \\
&\sigma_{j=odd}^+ &=& \exp\{-i \pi \big[\sum_{l<j}\left(A^\dag_l A_l + B^\dag_l B_l\right) + A^\dag_j A_j\big] \} B_j\nonumber
\end{aligned}
\end{equation}
\noindent with $\sigma_{j}^-=\left(\sigma_{j}^+\right)^{\dagger}$. We can now move to the Fourier space according to the definition 
\begin{SMEquations}
    A_j = \frac{1}{\sqrt{\mathcal{N}}}\sum_{q \in \Gamma} e^{i \frac{2\pi}{\mathcal{N}}q j} A_q
    \label{315}
\end{SMEquations}
\noindent and with an analogous expression for $B_{j}$, where $q$ runs within the set $\Gamma = \{\frac{1}{2},\frac{3}{2},\frac{5}{2}, ... , \mathcal{N} - \frac{1}{2} \}$. Due to the above mapping, the Hamiltonian in Eq.~\eqref{HB_Ladder} takes the form

\begin{SMEquations}
H_B = \frac{J}{2} \sum_{q \in \Gamma}
\textbf{A}_q^\dag
\mathcal{H}_{q}
\textbf{A}_q, \label{H320}
\end{SMEquations} 
where we have introduced the short notations $\textbf{A}^{\dagger}_{q}=\left(A_q^\dag, B_q^\dag, A_{\mathcal{N} - q}, B_{\mathcal{N} - q}\right)$ and

\begin{SMEquations}
\mathcal{H}_q = \begin{pmatrix}
0 & Z_{q} & 0 & W_{q} \\
Z_{q}^* & 0 & -W_{q}^* & 0 \\
0 & -W_{q} & 0 & -Z_{q} \\
-W_{q}^* & 0 & -Z_{q}^* & 0 
\end{pmatrix},  
\end{SMEquations}
with
\begin{SMEquations}
    \begin{aligned}
        &Z_{q} = -\left[(1+\delta) + (1-\delta) e^{-i \frac{2\pi}{\mathcal{N}}q}\right] \\
        &W_{q} = -\gamma\left[(1+\delta) - (1-\delta) e^{-i\frac{2\pi}{\mathcal{N}}q}\right].
    \end{aligned}
\end{SMEquations}
Considering now the unitary matrix $\mathcal{U}_{q}$ having the eigenvectors of $H_B$ as columns, we can write Eq.~\eqref{H320} as
\begin{SMEquations}
H_B = \frac{J}{2} \sum_{q \in \Gamma}
(\textbf{A}_q^\dag \mathcal{U}_{q})( \mathcal{U}_{q}^\dag 
\mathcal{H}_{q} \mathcal{U}_{q})( \mathcal{U}_{q}^\dag
\textbf{A}_q) = \frac{J}{2} \sum_{q \in \Gamma}
\textbf{a}_q^\dag
\mathcal{\Tilde{H}}_{q}
\textbf{a}_q, \label{SM7}
\end{SMEquations} 
where we introduced $\textbf{a}_q^\dag = \textbf{A}_q^\dag \mathcal{U}_{q} = \left(a_q^\dag, b_q^\dag, a_{\mathcal{N} - q}, b_{\mathcal{N} - q}\right)$ and
\begin{equation*}
    \mathcal{\Tilde{H}}_{q} = \mathcal{U}_{q}^\dag 
\mathcal{H}_{q} \mathcal{U}_{q} = \begin{pmatrix}
    \omega_{1,q} & 0 & 0 & 0 \\
    0 & \omega_{2,q} & 0 & 0 \\
    0 & 0 & - \omega_{1,q} & 0 \\
    0 & 0 & 0 & - \omega_{2,q}
\end{pmatrix}
\end{equation*}
The explicit form of \eqref{SM7}, assuming $J = 1$ as overall energy scale, reads
\begin{equation*}
    H_B = \sum_{q \in \Gamma} \left[\omega_{1,q} \left(a_q^\dag a_q - \frac{1}{2}\right) + \omega_{2,q} \left(b_q^\dag b_q - \frac{1}{2}\right)\right]
\end{equation*}
as reported in Eq.~(2) of the main text. 

\vspace{1cm}

\begin{center}
    \textbf{SM2: Numerical analysis of $\mathcal{M}_q$ and fermionic occupation numbers}
\end{center}
In Eq.~(7) of the main text we report the energy stored in the QB for a given time $t$ as
\begin{SMEquations}
    \Delta E(t) =  \sum_{q \in \Gamma} \sum_{s_1}\omega_{s_1,q} n_{s_1,q}(t) \label{E_n_q}
\end{SMEquations}
where 
\begin{SMEquations}
\begin{aligned}
n_{s_1,q}(t) = \sum_{s,s_2,s_3 = 1,2}2\bigg\{&\mathcal{M}_{s+2,s_{1}; q}\mathcal{M}^*_{s_{2}+2,s_{1}; q}\mathcal{M}^*_{s+2,s_3 + 2; q}\mathcal{M}_{s_{2}+2,s_3 + 2;q}~ \cos\left[\left(\omega'_{s,q} - \omega'_{s_{2},q}\right)t\right] + \\ &+ \mathcal{M}_{s+2,s_{1};q}\mathcal{M}^*_{s_{2},s_{1};q}\mathcal{M}^*_{s+2,s_3 + 2;q}\mathcal{M}_{s_{2},s_3 + 2;q}~ \cos\left[\left(\omega'_{s,q} + \omega'_{s_{2},q}\right)t\right] \bigg\} \label{n_occ}
\end{aligned}
\end{SMEquations}
is defined as $\langle \psi(t)|a^\dag_q a_q|\psi(t) \rangle$ if $s_1 = 1$ and $\langle \psi(t)|b^\dag_q b_q|\psi(t) \rangle$ if $s_1 = 2$. Although finding a simple analytical expression for $\mathcal{M}_q$ is challenging, a numerical analysis can be conducted in order to simplify Eq.~\eqref{E_n_q}. After fixing all the internal parameters of the system in the range reported in the main text, the terms in the sum that contribute significantly (i.e., those with an absolute value greater than $10^{-10}$, assumed as the tolerance for the numerical errors) are those with $s = s_1 = s_2 = s_3$, so we can write Eq.~\eqref{E_n_q} explicitly as:
\begin{SMEquations}
\begin{aligned}
    \Delta E(t) \simeq \sum_{q \in \Gamma} &\{\omega_{1,q} \left[2 ~|\mathcal{M}_{3,1; q}|^2 |\mathcal{M}_{3,3; q}|^2 + 2 \mathcal{M}_{3,1; q}\mathcal{M}^*_{1,1; q}\mathcal{M}^*_{3,3; q}\mathcal{M}_{1,3;q} ~ \cos(2\omega'_1 t)\right] + \\ &+ \omega_{2,q} \left[2 ~|\mathcal{M}_{4,2; q}|^2 |\mathcal{M}_{4,4; q}|^2 + 2 \mathcal{M}_{4,2; q}\mathcal{M}^*_{2,2; q}\mathcal{M}^*_{4,4; q}\mathcal{M}_{2,4;q} ~ \cos(2\omega'_2 t)\right]\}
\end{aligned}
\end{SMEquations}
so that 
\begin{equation*}
n_{1,q}(t) = 2 ~|\mathcal{M}_{3,1; q}|^2 |\mathcal{M}_{3,3; q}|^2 + 2 \mathcal{M}_{3,1; q}\mathcal{M}^*_{1,1; q}\mathcal{M}^*_{3,3; q}\mathcal{M}_{1,3;q} ~ \cos(2\omega'_1 t) 
\end{equation*} and 
\begin{equation*}
n_{2,q}(t) = 2 ~|\mathcal{M}_{4,2; q}|^2 |\mathcal{M}_{4,4; q}|^2 + 2 \mathcal{M}_{4,2; q}\mathcal{M}^*_{2,2; q}\mathcal{M}^*_{4,4; q}\mathcal{M}_{2,4;q} ~ \cos(2\omega'_2 t). 
\end{equation*}
From this result, it can be observed that the term which is proportional to $\cos\left[\left(\omega'_{s,q} - \omega'_{s_{2},q}\right)t\right]$ in Eq.~\eqref{n_occ} is indeed constant and represents the thermodynamic regime discussed in the main text, while all the temporal oscillations are encoded in the term proportional to $\cos\left[\left(\omega'_{s,q} + \omega'_{s_{2},q}\right)t\right]$. It is also worth noting that, in the parameter region we are considering, only the lowest energy band, namely the one associated to the $b_q$ fermions, gets indeed filled. Therefore, the expression for the energy stored at a given time $t$ can be further simplified as follows
\begin{SMEquations}
    \Delta E(t) \simeq \sum_{q \in \Gamma}\omega_{2,q} n_{2,q}(t).
\end{SMEquations}

To achieve a more in-depth understanding of the presence of peaks corresponding to phase transitions in the asymptotic and first revival regimes, the behavior of $n_{2,q}(\tau_\infty)$ and $n_{2,q}(\tau_R)$ as a function of $k = \frac{2\pi}{\mathcal{N}}q$ for these values of $\delta_0$ has been studied.  In the following we will report the situation corresponding to Fig.~3 of the main text.

\begin{figure}[ht]
    \centering
    \begin{subfigure}[b]{0.5\textwidth}
        \centering
        \includegraphics[width=\textwidth]{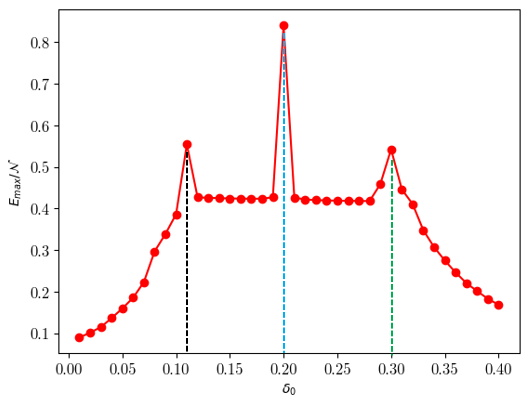}
        \caption{}
        \label{fig:image1}
    \end{subfigure}
    
    
    \begin{subfigure}[b]{0.3\textwidth}
        \centering
        \includegraphics[width=\textwidth]{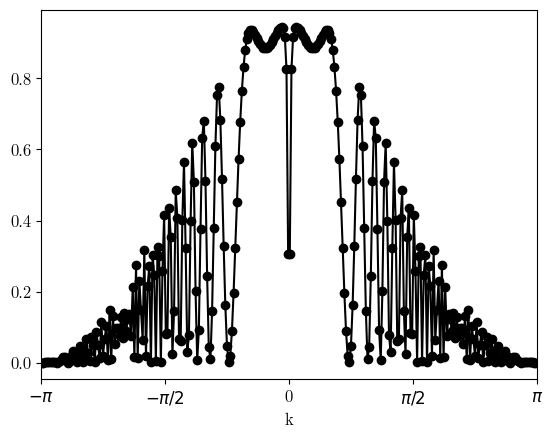}
        \caption{}
        \label{fig:image2}
    \end{subfigure}
    \hspace{0.1cm}
    \begin{subfigure}[b]{0.3\textwidth}
        \centering
        \includegraphics[width=\textwidth]{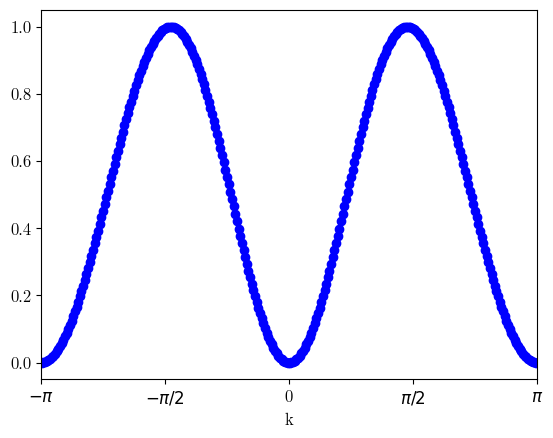}
        \caption{}
        \label{fig:image3}
    \end{subfigure}
    \hspace{0.1cm}
    \begin{subfigure}[b]{0.3\textwidth}
        \centering
        \includegraphics[width=\textwidth]{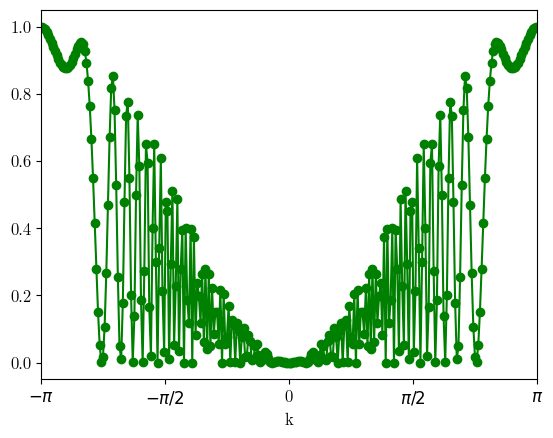}
        \caption{}
        \label{fig:image4}
    \end{subfigure}
    
    \caption{Plot of $E_{max}/\mathcal{N}$ as a function of $\delta_0$ for $\gamma = 1.1$, $\delta_1 = 0.8$ and $\mathcal{N} = 300$ at time $t = \tau_R$ \textbf{(a)} and plot of $n_{2,q}(\tau_R)$ as a function of $k = \frac{2\pi}{\mathcal{N}}q$ for $\delta_0 = 0.11$ \textbf{(b)}, $\delta_0 = 0.2$ \textbf{(c)} and $\delta_0 = 0.3$ \textbf{(d)} in correspondence of the dashed vertical lines of panel (a).}
    \label{Complete_1.1_0.8}
\end{figure}

\begin{figure}[H]
    \centering
    \begin{subfigure}[b]{0.5\textwidth}
        \centering
        \includegraphics[width=\textwidth]{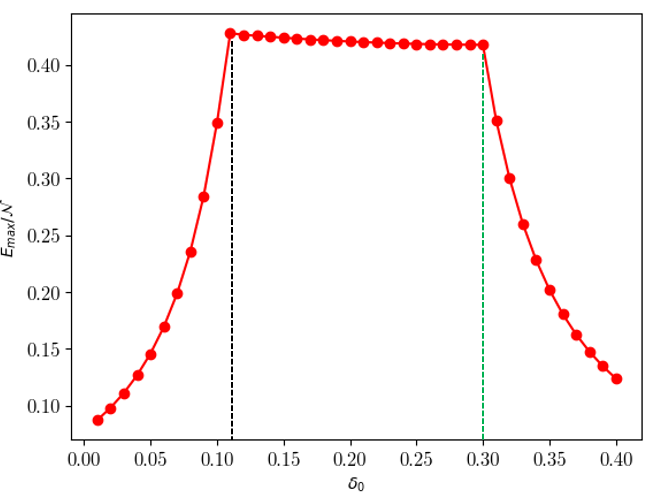}
        \caption{}
        \label{fig:image1}
    \end{subfigure}
    
    
    \begin{subfigure}[b]{0.3\textwidth}
        \centering
        \includegraphics[width=\textwidth]{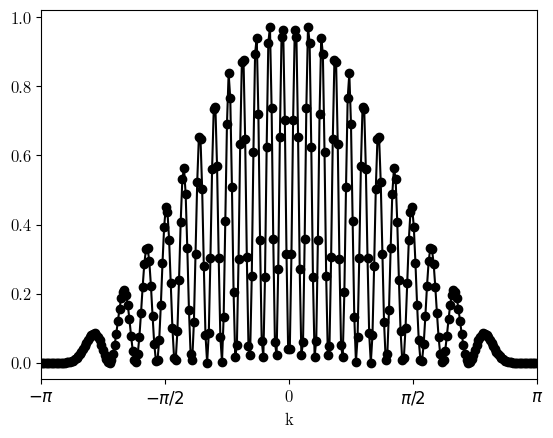}
        \caption{}
        \label{fig:image2}
    \end{subfigure}
    \hspace{0.1cm}
    \begin{subfigure}[b]{0.3\textwidth}
        \centering
        \includegraphics[width=\textwidth]{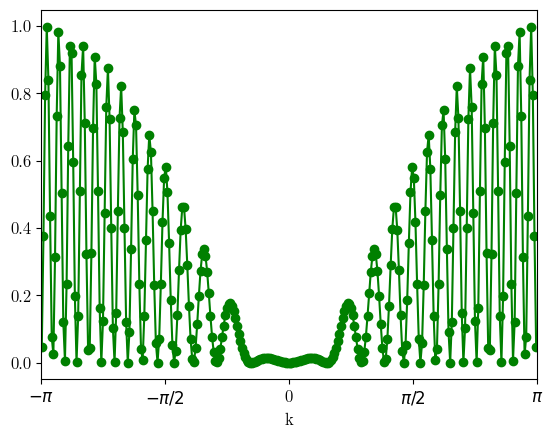}
        \caption{}
        \label{fig:image3}
    \end{subfigure}
    
    \caption{Plot of $E_{max}/\mathcal{N}$ as a function of $\delta_0$ for $\gamma = 1.1$, $\delta_1 = 0.8$ and $\mathcal{N} = 300$ at time $t = \tau_\infty$ \textbf{(a)} and plot of $n_{2,q}(\tau_\infty)$ as a function of $k = \frac{2\pi}{\mathcal{N}}q$ for $\delta_0 = 0.11$ \textbf{(b)} and $\delta_0 = 0.3$ \textbf{(c)} in correspondence of the dashed vertical lines of panel (a).} \label{Asympt_1.1_0.8}
\end{figure}

As shown in Fig.~\ref{Complete_1.1_0.8} and \ref{Asympt_1.1_0.8}, the emergence of these peaks is justified by the high number of fermions $b_q$ (approaching $1$) occupying the region around the value of $k$ at which the transition occurs ($k = 0$ for $\delta_0 = 0.11$ and $k = \pm \pi$ for $\delta_0 = 0.3$), while in the case of flat bands, where $\delta_0 + \delta_1 = 1$,  the fermions predominantly occupy the region around $k = \frac{\pi}{2}$. It is worth to note that in correspondence of the phase transitions the fermion occupation numbers are characterized by beats associated to various competing oscillatory contributions. 


\begin{figure}[H]
    \centering
    \begin{subfigure}[b]{0.3\textwidth}
        \centering
        \includegraphics[width=\textwidth]{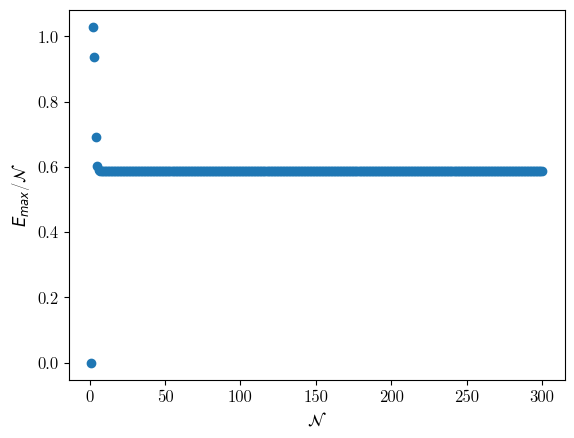}
        \caption{}
        \label{Energy_N_s}
    \end{subfigure}
    \begin{subfigure}[b]{0.3\textwidth}
        \centering
        \includegraphics[width=\textwidth]{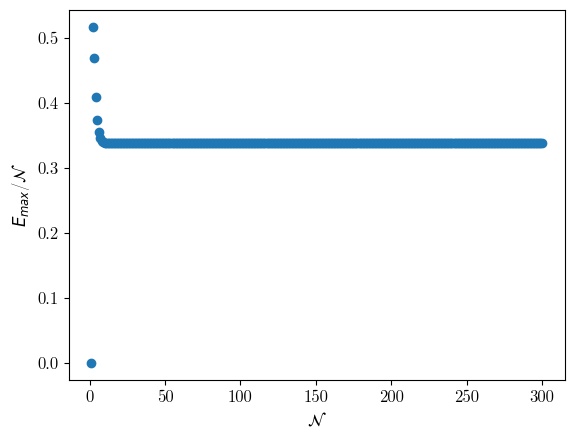}
        \caption{}
        \label{Energy_N_r}
    \end{subfigure}
    \hspace{0.1cm}
    \begin{subfigure}[b]{0.3\textwidth}
        \centering
        \includegraphics[width=\textwidth]{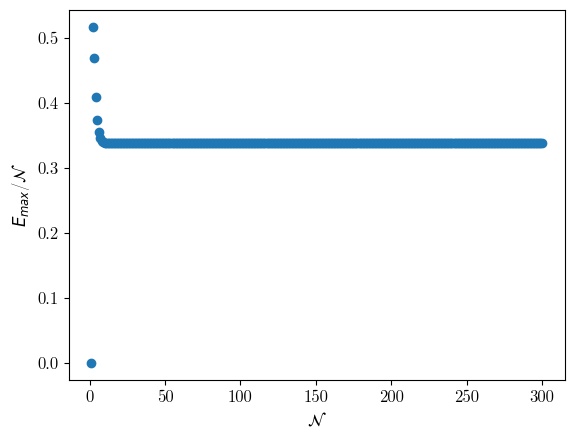}
        \caption{}
        \label{Energy_N_a}
    \end{subfigure}
    \caption{Plot of $E_{max}(\tau_s)/\mathcal{N}$ \textbf{(a)}, $E_{max}(\tau_r)/\mathcal{N}$ \textbf{(b)} and $E_{max}(\tau_\infty)/\mathcal{N}$ \textbf{(c)} as a function of $\mathcal{N}$ for $\gamma = 1.25, \delta_0 = 0.3, \delta_1 = 0.6$.} \label{Energy_respect_N}
\end{figure}

\begin{figure}[H]
    \centering
    \includegraphics[width=0.5\textwidth]{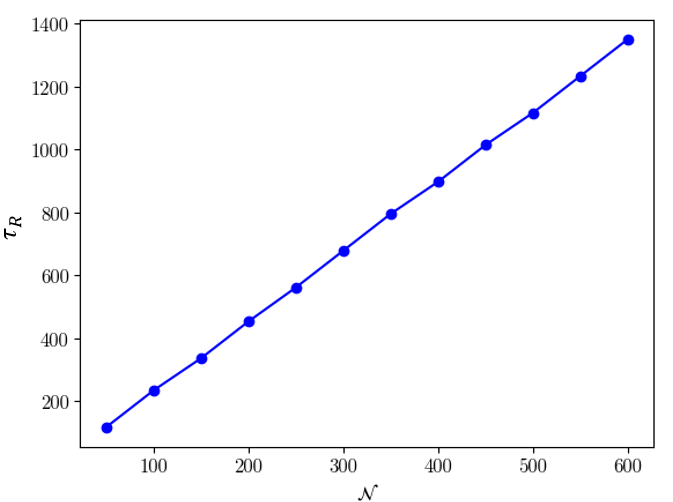}
    \caption{Plot of $\tau_R$ as a function of $\mathcal{N}$ for $\gamma = 1.25, \delta_0 = 0.3, \delta_1 = 0.6$.}
    \label{tau_su_n}
\end{figure}

\begin{center}
    \textbf{SM3: Scaling of energy and time revival}
\end{center}
In this section, we show the behavior of the maximum energy capacity of the quantum battery, evaluated in the three different regimes, as a function of the number of dimers in the chain, to highlight the fact that it is linear with $\mathcal{N}$ (Figs. \ref{Energy_N_s}, \ref{Energy_N_r} and \ref{Energy_N_a}). Additionally, in Fig.~\ref{tau_su_n}, it can be seen that the time $\tau_R$ at which the first revival occurs also scales with $\mathcal{N}$ as the length of the chain increases. This is in agreement with what reported in \cite{rossini2020dynamics}, thus justifying the absence of a revival in the thermodynamic limit.

\begin{center}
    \textbf{SM4: Calculation for Ising model}
\end{center}

In this section we discuss the explicit analytical computation of $\Delta E (t) = \langle \Psi(t)|H_{B}|\Psi(t)\rangle - \langle \Psi(0)|H_{B}|\Psi(0)\rangle$ for the Ising model in presence of an external field
\begin{SMEquations}
    H = \frac{1}{2} \sum_{j=1}^N \left[\sigma_j^x \sigma_{j+1}^x+h \sigma_j^z\right]. \label{XY_Ham}
\end{SMEquations}
\noindent Using the same techniques already described in section \textbf{SM1}, a Jordan-Wigner mapping can be defined between spin operators $\sigma^{\pm}$ and fermionic operators $\psi_q$ leading to (see Ref.\cite{franchini2017introduction} for more details)
\begin{SMEquations}
H = \frac{1}{2} \sum_{q \in \Gamma}
\begin{pmatrix}
\psi_q^\dag & \psi_{N - q}
\end{pmatrix}
\begin{pmatrix}
h - \cos\left(\frac{2\pi}{N}q\right) & -\sin\left(\frac{2\pi}{N}q\right) \\
-\sin\left(\frac{2\pi}{N}q\right) & \cos\left(\frac{2\pi}{N}q\right) - h
\end{pmatrix}
\begin{pmatrix}
\psi_q \\ \psi_{N - q}^\dag
\end{pmatrix}.\label{H_matrix}
\end{SMEquations}
Now, by means of the following Bogoliubov transformations
\begin{SMEquations}
\begin{pmatrix}
\psi_q \\ \psi_{N - q}^\dag
\end{pmatrix} = 
\begin{pmatrix}
\cos \theta_q & \sin \theta_q \\
-\sin \theta_q & \cos \theta_q
\end{pmatrix}
\begin{pmatrix}
a_q \\ a_{N - q}^\dag
\end{pmatrix}, \label{bog}
\end{SMEquations}
where $\theta_q$ satisfies
\begin{SMEquations}
e^{i 2\theta_q} = \frac{h - \cos\left(\frac{2\pi}{N}q\right) + i \sin\left(\frac{2\pi}{N}q\right)}{\sqrt{1 + h^2 - 2h \cos(\frac{2\pi}{N}q)}}, \label{ConditionTheta}
\end{SMEquations}
we obtain the diagonal form of the Hamiltonian
\begin{SMEquations}
    H = \sum_{q \in \Gamma} \epsilon_q \left[a_q^\dag a_q - \frac{1}{2}\right] \label{DiagHam}
\end{SMEquations}
where
\begin{SMEquations}
    \epsilon_q = \sqrt{1 + h^2 - 2h \cos(\frac{2\pi}{N}q)}.
\end{SMEquations}

Let's consider the most general case, where we start with the initial hamiltonian $H(h_i)$ written in the form of Eq. \eqref{DiagHam} with the dynamics starting in the ground state $|\psi(0) \rangle$, vacuum of the $a_q$ fermions, and we let the state evolve according to  
\begin{SMEquations}
    H (h_f) = \sum_{q \in \Gamma} \omega_q \left[b_q^\dag b_q - \frac{1}{2}\right]
\end{SMEquations}
diagonal in terms of the new fermions $b_q$.

In analogy with what done in the main text, the goal is to compute (putting $\hbar = 1$)
\begin{SMEquations}
\begin{aligned}
    \Delta E (t) &= \langle \Psi(t)|H(h_i)|\Psi(t)\rangle - \langle \Psi(0)|H(h_i)|\Psi(0)\rangle = \\ &= \langle \Psi(0)|e^{i H(h_f) t} H(h_i) e^{-i H(h_f) t}|\Psi(0)\rangle + \frac{1}{2} \sum_{q \in \Gamma} \epsilon_q. \label{DeltaEFull}
\end{aligned}
\end{SMEquations}

To compute the bracket, we first recall the relation between $a_q$ fermions and $\psi_q$ fermions from Eq. \eqref{bog}
\begin{SMEquations}
    a_q = \cos \theta_q^{(i)} \psi_q - \sin \theta_q^{(i)} \psi^\dag_{N-q} \label{TrasfA}
\end{SMEquations}
where $\theta_q^{(i)} = \theta_q(h_i)$. Notice that in this case a rotation can be used due to the fact that here the mapping from spin into fermions is not characterized by the presence of superconducting-like coupling terms. 

\begin{figure}[H]
    \centering
    \includegraphics[width=0.5\textwidth]{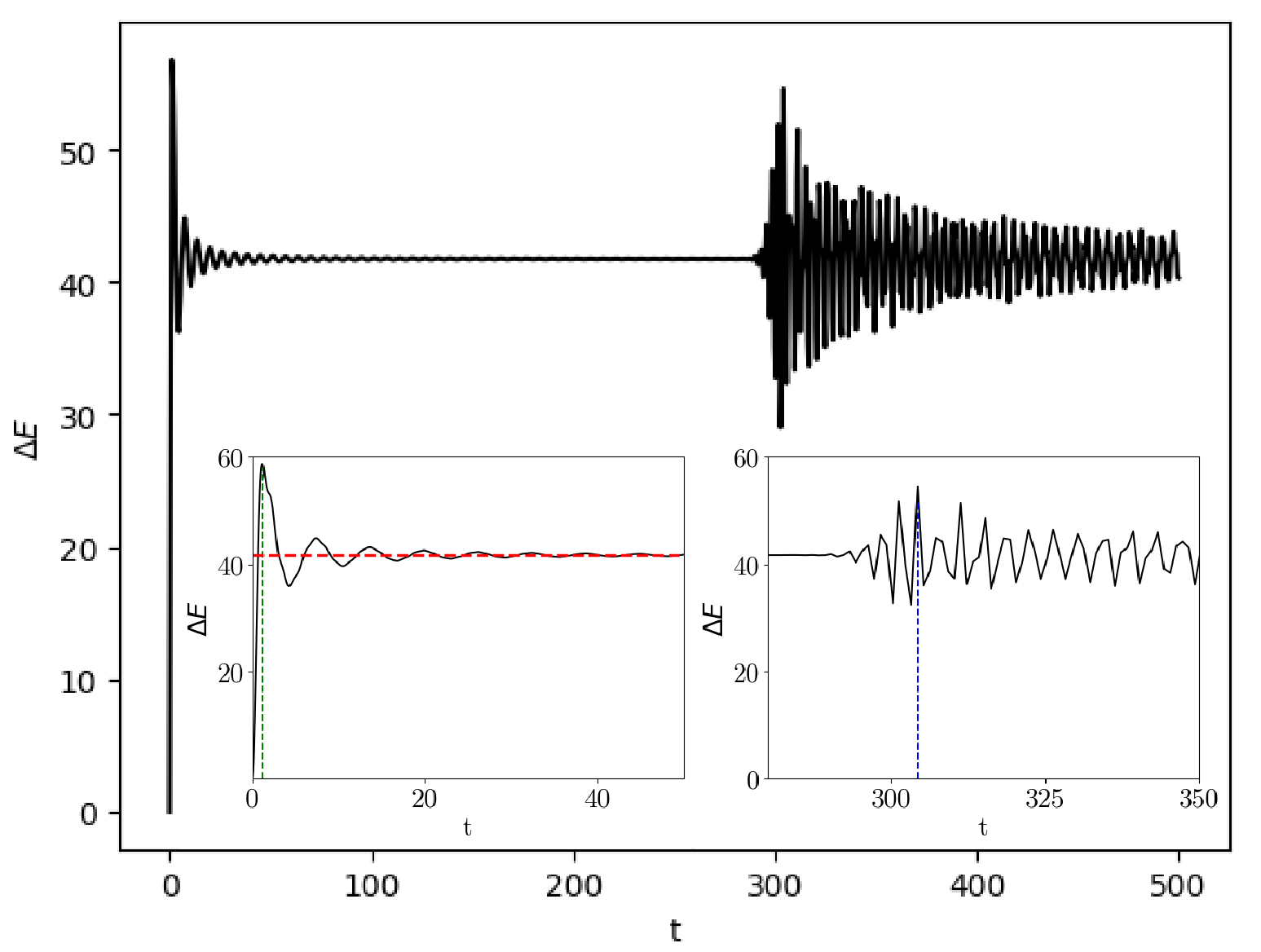}
    \caption{Energy $\Delta E$ stored in the QB as a function of time for $h_0 = 0.8$, $h_1 = 0.7$ and $N = 600$. Left inset: zoom for $t \in \left[0, 50\right]$. The vertical green dotted line indicates the time at which the first maximum of the energy stored is achieved, namely $\tau_s$, while the horizontal red dotted line indicates the asymptotic limit of the maximum energy stored per dimer. Right inset: zoom for $t \in \left[280, 350\right]$. The vertical blue dotted line indicates the time at which the maximum of the energy stored during the recurrence is achieved, namely $\tau_R$.}
    \label{IsingTime}
\end{figure}

Analogously, we can write the relations for $b_q$ fermions
\begin{SMEquations}
\begin{aligned}
    &\psi_q = \cos \theta_q^{(f)} b_q + \sin \theta_q^{(f)} b^\dag_{N-q} \\
    & \psi^\dag_{N-q} = \cos \theta_q^{(f)} b^\dag_{N-q} - \sin \theta_q^{(f)} b_q \label{TrasfB}
\end{aligned}
\end{SMEquations}
where $\theta_q^{(f)} = \theta_q(h_f)$ and the property $\theta_{N-q}^{(f)} = -\theta_q^{(f)}$ has been used. Substituting Eqs. \eqref{TrasfB} into Eq. \eqref{TrasfA} leads to
\begin{SMEquations}
    a_q = \cos \theta_q^{(i)} \left[\cos \theta_q^{(f)} b_q + \sin \theta_q^{(f)} b^\dag_{N-q}\right] - \sin \theta_q^{(i)} \left[\cos \theta_q^{(f)} b^\dag_{N-q} - \sin \theta_q^{(f)} b_q\right].
\end{SMEquations}
We can now evolve with $H(h_f)$ using the relation
\begin{SMEquations}
    b_q(t) = e^{-i \omega_q t} b_q
\end{SMEquations}
to obtain
\begin{SMEquations}
\begin{aligned}
    a_q (t) &= \cos \theta_q^{(i)} \left[\cos \theta_q^{(f)} e^{-i \omega_q t} b_q + \sin \theta_q^{(f)} e^{i \omega_q t} b^\dag_{N-q}\right] - \sin \theta_q^{(i)} \left[\cos \theta_q^{(f)} e^{i \omega_q t} b^\dag_{N-q} - \sin \theta_q^{(f)} e^{-i \omega_q t} b_q\right] = \\ &= e^{-i \omega_q t} b_q \cos\left(\theta_q^{(i)} - \theta_q^{(f)}\right) - e^{i \omega_q t} b^\dag_{N-q} \sin\left(\theta_q^{(i)} - \theta_q^{(f)}\right). \label{Aqt}
\end{aligned}
\end{SMEquations}
Finally, we write Eq.\eqref{Aqt} in terms of $a_q$ fermions substituting
\begin{SMEquations}
\begin{aligned}
    &\psi_q = \cos \theta_q^{(i)} a_q + \sin \theta_q^{(i)} a^\dag_{N-q} \\
    & \psi^\dag_{N-q} = \cos \theta_q^{(i)} a^\dag_{N-q} - \sin \theta_q^{(i)} a_q 
\end{aligned}
\end{SMEquations}
into
\begin{SMEquations}
    b_q = \cos \theta_q^{(f)} \psi_q - \sin \theta_q^{(f)} \psi^\dag_{N-q}
\end{SMEquations}
obtaining
\begin{SMEquations}
\begin{aligned}
    &b_q = \cos\left(\theta_q^{(i)} - \theta_q^{(f)}\right) a_q + \sin\left(\theta_q^{(i)} - \theta_q^{(f)}\right) a^\dag_{N-q} \\
    & b^\dag_{N-q} = \cos\left(\theta_q^{(i)} - \theta_q^{(f)}\right) a^\dag_{N-q} - \sin\left(\theta_q^{(i)} - \theta_q^{(f)}\right) a_q.
\end{aligned}
\end{SMEquations}
After implementing these new transformations in Eq.\eqref{Aqt}, the result is
\begin{SMEquations}
    a_q (t) = \alpha(t)~ a_q + \beta(t)~ a^\dag_{N-q}
\end{SMEquations}
where
\begin{SMEquations}
\begin{aligned}
    &\alpha (t) = e^{-i \omega_q t} \cos^2\left(\theta_q^{(i)} - \theta_q^{(f)}\right) + e^{i \omega_q t} \sin^2 \left(\theta_q^{(i)} - \theta_q^{(f)}\right) \\
    & \beta (t) = -2i \sin(\omega_q t)\sin\left(\theta_q^{(i)} - \theta_q^{(f)}\right)\cos\left(\theta_q^{(i)} - \theta_q^{(f)}\right).
\end{aligned}
\end{SMEquations}
Since we are interested in computing
\begin{SMEquations}
    \langle \psi(0) | a_q^\dag(t) a_q(t) |\psi(0) \rangle
\end{SMEquations}
and $a_q |\psi(0) \rangle = 0$ by definition, the only term of the product $a_q^\dag(t) a_q(t)$ that gives a non-zero contribution is $a_{N-q} a_{N-q}^\dag$, so
\begin{SMEquations}
    \langle \psi(0) | a_q^\dag(t) a_q(t) |\psi(0) \rangle = \langle \psi(0) | a_{N-q} a_{N-q}^\dag |\psi(0) \rangle = |\beta(t)|^2.
\end{SMEquations}
Using this result in Eq.\eqref{DeltaEFull} leads to
\begin{SMEquations}
    \Delta E (t) = \left(\sum_{q \in \Gamma}|\beta(t)|^2  - \frac{1}{2} \sum_{q \in \Gamma} \epsilon_q\right)  + \frac{1}{2} \sum_{q \in \Gamma} \epsilon_q = \sum_{q \in \Gamma}|\beta(t)|^2 = \sum_{q \in \Gamma}\left[4\sin^2(\omega_q t)\sin^2\left(\theta_q^{(i)} - \theta_q^{(f)}\right)\cos^2\left(\theta_q^{(i)} - \theta_q^{(f)}\right)\right].
\end{SMEquations}
After some algebra, we can write the result as
\begin{SMEquations}
    \Delta E (t) = \sum_{q \in \Gamma}\bigg\{\frac{\epsilon_q}{2} \left[1 - \cos(2 \omega_q t)\right]\left[\sin(2\theta_q^{(i)})\cos(2\theta_q^{(f)}) - \cos(2\theta_q^{(i)})\sin(2\theta_q^{(f)})\right]^2\bigg\}.
\end{SMEquations}
Finally, we recall Eq. \eqref{ConditionTheta} which gives the following set of relations:
\begin{SMEquations}
\begin{aligned}
    &\sin(2\theta_q^{(i)}) = \Im[e^{i 2\theta_q^{(i)}}] = \frac{\sin\left({\frac{2\pi}{N}q}\right)}{\epsilon_q} \\
    &\cos(2\theta_q^{(i)}) = \Re[e^{i 2\theta_q^{(i)}}] = \frac{h_i - \cos\left({\frac{2\pi}{N}q}\right)}{\epsilon_q} \label{34}
\end{aligned}
\end{SMEquations}
and
\begin{SMEquations}
\begin{aligned}
    &\sin(2\theta_q^{(f)}) = \Im[e^{i 2\theta_q^{(f)}}] = \frac{\sin\left({\frac{2\pi}{N}q}\right)}{\omega_q} \\
    &\cos(2\theta_q^{(f)}) = \Re[e^{i 2\theta_q^{(f)}}] = \frac{h_f - \cos\left({\frac{2\pi}{N}q}\right)}{\omega_q} \label{35}
\end{aligned}
\end{SMEquations}
With Eqs. \eqref{34} and Eqs. \eqref{35} we can write the final form of the energy stored at time $t$ as
\begin{SMEquations}
    \Delta E (t) = \sum_{q \in \Gamma}\bigg\{\frac{1}{2\epsilon_q \omega^2_q} \sin^2\left({\frac{2\pi}{N}q}\right)\left(h_f - h_i \right)^2 \left[1 - \cos(2 \omega_q t)\right]\bigg\}.
\end{SMEquations}
To maintain the notation of the main text, we set $h_i = h_0$ and $h_f = h_0 + h_1$ obtaining
\begin{SMEquations}
    \Delta E (t) = \sum_{q \in \Gamma}\bigg\{\frac{h_1^2}{2\epsilon_q \omega^2_q} \sin^2\left({\frac{2\pi}{N}q}\right)\left[1 - \cos(2 \omega_q t)\right]\bigg\}.
\end{SMEquations}
For the Ising model as well, we can plot the stored energy as a function of time, as shown in Fig.\ref{IsingTime}, observing a trend similar to that in Fig.2 of the main text.
